\documentclass{iau_FM}
\usepackage{graphicx}
\usepackage{subfig}
\usepackage{sidecap}

\usepackage{floatrow}

\title[ON 231 in outburst state] 
{ {\it Swift} and {\it XMM-Newton} monitoring of the ISP blazar ON 231 in outburst state} 

\author[Nibedita Kalita \& Utane Sawangwit ]   
{Nibedita\ Kalita$^{1}$ \and Utane\ Sawangwit$^{1}$}

\affiliation{$^1$National Astronomical Research Institute of Thailand (NARIT), Chiang Mai, 50180, Thailand \\ email:{\tt nibedita@narit.or.th}}
 
\pubyear{2018}
\jname{Astronomy in Focus, Volume 1} 
\editors{Volker Beckmann \& Claudio Ricci, eds.}
\begin{document}
\maketitle

\begin{abstract}
We present a detailed temporal and spectral study of the intermediate type blazar, ON 231 with observations taken by {\it Swift} and {\it XMM-Newton} satellites during an outburst phase in June $2008$. The source shows multiple X-ray flares in that period when the flux amplitude varies in the range of $27-38\%$. The X-ray spectra are well fitted with a broken power-law model indicating the presence of both synchrotron and inverse Compton components. We find that the source shows strong and variable emission in the soft energy bands (below $3-4$ keV), where the hard band emission is weak and stable. All the soft X-ray bands show correlated variability with zero lag, while a soft lag of -600s between the $0.3-0.5$ and $4-10$ keV bands is observed with DCF analysis. A time-resolved spectral study of the flares gives a positive relation between the total fluxes and the break energies of the two emission components.


\end{abstract}

\firstsection 

\section{Introduction}
 
ON 231, also known as W Comae ({\it z} = 0.102) was the 1st intermediate type synchrotron peaked (ISP) blazar detected in TeV $\gamma-ray$ range by {\it VERITAS} in March, 2008 when the source went to an outburst phase (\cite[Acciari et al. 2008]{Acciari_etal08}). The source entered into a second outburst state in June in the same year, which triggered a multi-frequency campaign to observe the source in different wavelengths. This time the gamma-ray flux was higher than the previous flare by a factor of 3. The observations that we have used in this work were taken as part of the multi-wavelength campaign. Assuming that the photons are emitted due to shocks in the jet during the outburst state, \cite[Sorcia et al. (2014)]{Sorcia_etal14} estimated the polarization degree, $\sim$ 33.3$\%$ and the jet angle, $\sim$ 2$^\circ$. They also reported that the optical gamma-ray emitting regions could be co-spatial. The source has not much studied in the X-ray bands as compared to other blazars (LSPs or HSPs). However, a previous study reported the presence of variable soft and stable hard X-ray emission in the source with {\it BeppoSAX} observations (\cite[Tagliaferri et al. 2000]{Tagliaferri_etal00}).

The motivation behind this work is to investigate thoroughly the variability nature of the ISP blazar in flaring state. It is important to note that the ISPs are the connecting link between the high synchrotron peaked (HSPs) and the low synchrotron peaked (LSPs) blazars, so their study might unfold interesting information to understand the blazar phenomenon as a whole.

\section{Results: Timing and spectral analysis}

{\it Swift} and {\it XMM-Newton} satellites monitored the blazar in consecutive periods between 7$-$18 June 2008. We extract all the background subtracted light curves in the energy range 0.3-10 keV using the data processing method described in \cite[Kalita et al. 2015]{Kalita_etal15} for {\it XMM-Newton}'s EPIC-pn observations and the online Build-XRT-products for {\it Swift-XRT} observations. The fractional rms variability amplitude (F$_{var}$), estimated using the method given in \cite[Edelson et al. (2002)]{Edelson_etal02} indicates high flux variation ($\sim$ 27$-$38$\%$) on timescales of hours with 100$\%$ duty cycle. 

In order to study the energy-dependent variability properties of the event, we divide the 0.3-10 keV LCs into different energy bands and find that the emission in soft bands is strong and more variable than that in hard bands. To search for correlated variation among these energy bands, we have used Discrete Co-correlation Function (DCF) given by \cite[Edelson et al. 1988]{Edelson88}. The result of DCF analysis is the same for all the soft bands where the peak appeared at DCF$=$1 with zero time lag. However, a soft lag of -600s is detected between the 0.3$-$0.5 and 4$-$10 keV bands.

\begin{figure*}      
\centering
\mbox{\subfloat{\includegraphics[scale=0.45]{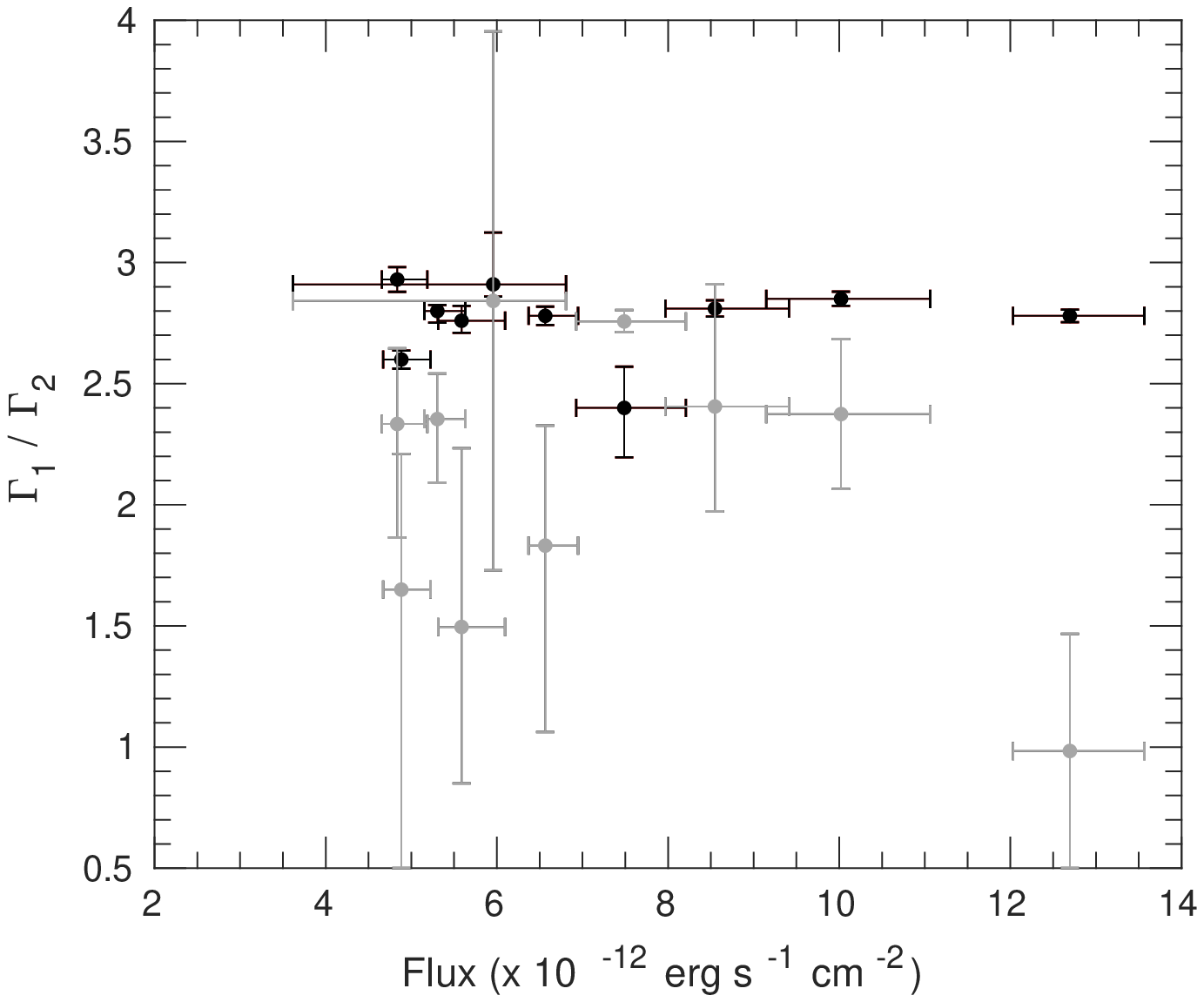}}\quad
\subfloat{\includegraphics[scale=0.45]{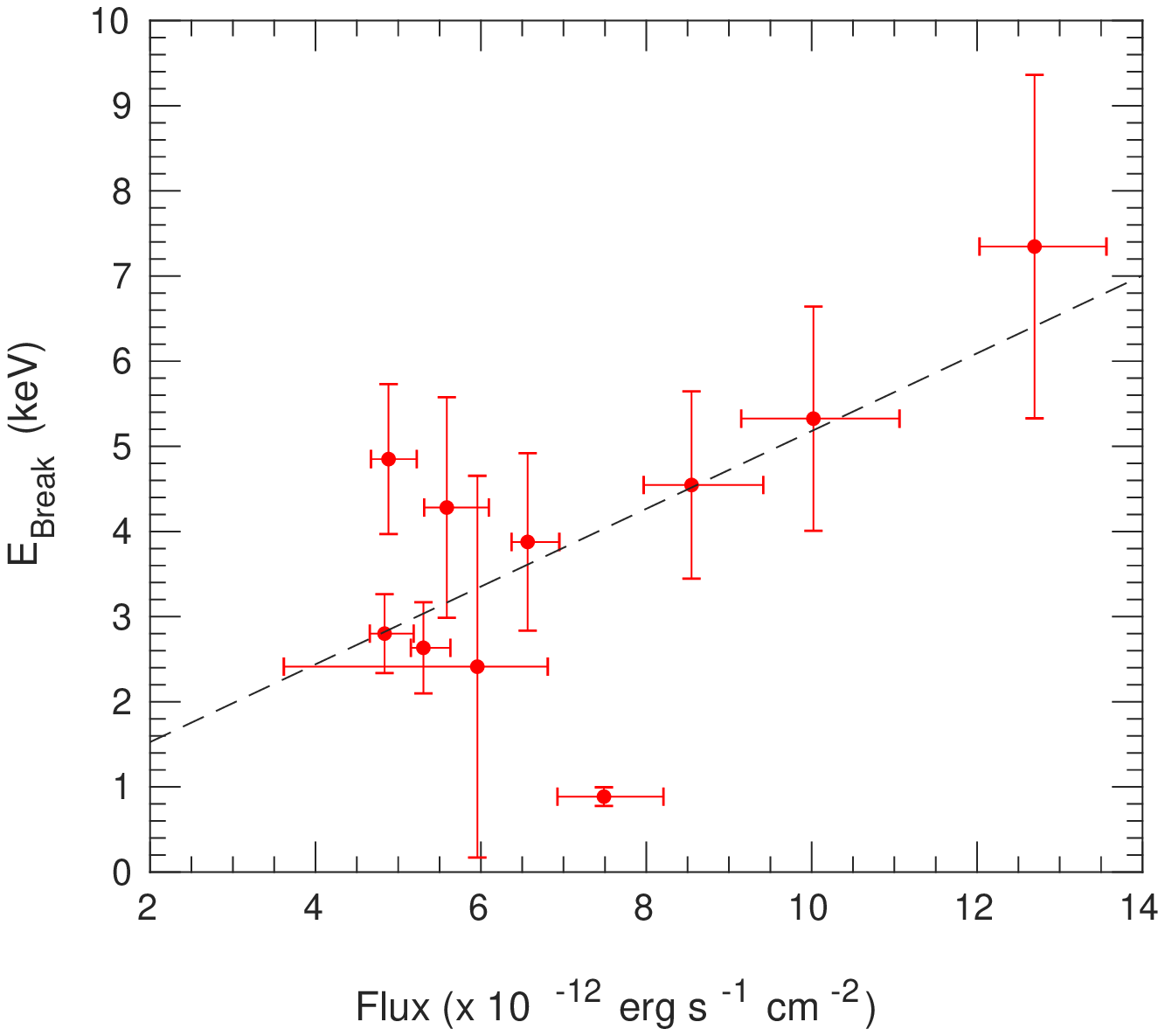} }}
\caption{The left plot shows the relationship between the two spectral indices and the 0.6-10 keV total fluxes derived from the BPL fit. The black and grey points represent $\Gamma_{1}$ and $\Gamma_{2}$, respectively. The plot for the break energies vs. the total fluxes is shown in the right panel of the figure.}
\label{fig1}
\end{figure*}

For all of the spectra when fitted with a simple power-law model with Galactic absorption due to the hydrogen column density, give positive residuals above 3-4 keV. This indicates that the spectra are intrinsically curved. So, we tried to fit the spectra with a log-parabolic model, but the residuals remain the same as in the previous case. However, a broken power-law (BPL) model gives an acceptable fit to all of the spectra with break energy (E$_{break}$) ranging between 2.6$-$5.4 keV and the two photon indices, $\Gamma_{1}$ and $\Gamma_{2}$ ranging from 2.74$\pm$0.01$-$3.03$\pm$0.1 and 1.74$\pm$0.35$-$2.4$\pm$0.15, respectively. To understand the evolution of the X-ray flares, we perform a time-resolved spectral analysis where all the spectra are split into several time intervals. We repeat spectral fits for each time episode with the best fit model and estimate the model parameters. The results of the time-resolved spectral fits are shown in the Fig.\,\ref{fig1}.

\end{document}